\documentclass[aps,pra,twocolumn,superscriptaddress,showpacs,floatfix]{revtex4-2}
\usepackage{hyperref}
\usepackage[british]{babel}
\usepackage[utf8]{inputenc}
\usepackage{amsmath}
\usepackage{amssymb}
\usepackage{mathtools}
\usepackage{braket}
\usepackage{graphicx}
\usepackage{xcolor}
\usepackage{dsfont}

\begin{document}
\title{Comparing quantum and classical finite state generators} 
\author{Prasenjit Deb}
\affiliation{Center for Quantum Engineering, Research, and Education, TCG CREST, Bidhan Nagar, Kolkata - 700091, India}
\author{Almut Beige}
\affiliation{The School of Physics and Astronomy, University of Leeds, Leeds LS2 9JT, United Kingdom}
\author{Lewis A. Clark}
\affiliation{Quantum Innovation Centre (Q.InC), Agency for Science Technology and Research (A*STAR), 4 Fusionopolis Way, Kinesis \#05-01, Singapore 138635, Republic of Singapore}

\date{\today}

\begin{abstract}
Bell-CHSH-like inequalities have been very successful in benchmarking {\it spatial} quantum correlations. However, as this paper illustrates, they are not necessarily sufficient for benchmarking {\it temporal} quantum correlations. To illustrate this, we parametrise classical and quantum stochastic finite state generators based on a single bit and a single qubit, respectively, and compare the temporal correlations of their output sequences using a Bell-CHSH-like inequality. We find that for sequential measurements by two observers, classical machines can exceed the Tsirelson bound of $2\sqrt{2}$. To present at least one example in which quantum machines outperform their classical counterparts and maintain correlations for longer, we consider general scrambling operations between consecutive measurements. Our analysis shines new light on the fundamental differences between quantum and classical finite state generators and can potentially be used to identify novel resources for quantum technology applications.
\end{abstract}

\maketitle

\section{Introduction} \label{sec:1}

Quantum entanglement, which can be used to obtain non-classical correlations in space-like separated scenarios \cite{aspect,EPR,hardy,bell,PR_box}, is seen by many as the most important resource for quantum information processing, including tasks such as quantum cryptography \cite{E91}, quantum teleportation \cite{tele}, quantum computing \cite{DiVincenzo} and dense coding \cite{densecoding}. However, in general, it is not easy to generate quantum entanglement \cite{EPR}. It is therefore worth noting that there are other quantum resources, like temporal quantum correlations. These correlations are easier to realise experimentally, since their generation only requires a series of generalised measurements on a single quantum system. For example, the BB84 quantum cryptography protocol \cite{BB84} does not require entanglement and relies instead on temporal quantum correlations due to subsequent measurements on the same photon. Nevertheless, its performance has many similarities with Ekert's protocol, which is entanglement-based \cite{E91}. Other applications of temporal quantum correlations can already be found in quantum metrology protocols and in other quantum technology schemes based on quantum feedback control \cite{Lewis1,Lewis2,Kawthar1}.

\begin{figure}[t] 
	\centering
	\includegraphics[width=.99\linewidth]{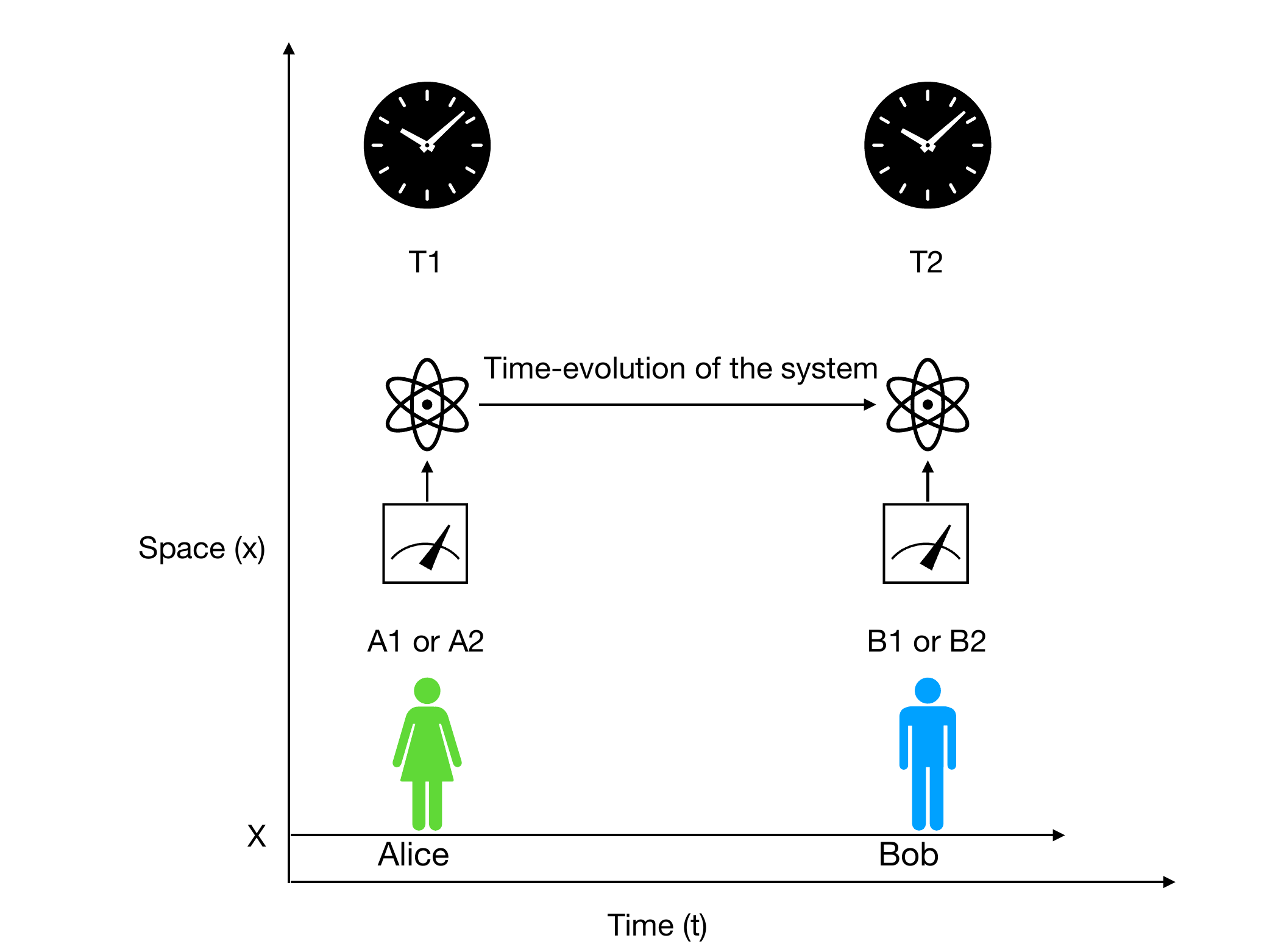}
	\caption{Schematic diagram of the scenario that we consider in this paper. In the quantum case, Alice performs a generalised measurement on a single qubit at time T1. Afterwards, she leaves the lab and lets Bob perform a subsequent measurement on the resulting quantum state at time T2. Here we are especially interested in the temporal correlations between Alice and Bob's measurement outcomes for randomly selected generalised measurements.
	}
	\label{figure1}
\end{figure}

In recent years, a lot of work has been done benchmarking spatial correlations, i.e.~the amount of entanglement, shared between distant qubits. One of the simplest scenarios in this context is the one introduced by Bell \cite{bell} and later considered by Clauser, Horne, Shimony and Holt \cite{chsh}.  Two parties, namely, \textit{Alice} and \textit{Bob}, each with a physical system of their own, conduct one of two possible dichotomic measurements.  Then, by comparing the correlations that they share and by checking for a violation of Bell-CHSH inequalities, they can determine whether or not their quantum systems have been entangled or not. It is known that classical models defined in terms of local hidden variable models \cite{kochen,mermin,bohm} obey these inequalities, whereas, if the two parties share a sufficient amount of entanglement, these inequalities are violated. The maximal quantum violation of the CHSH inequalities was found to be $2\sqrt{2}$ \cite{chsh,bell}, generally known as the \textit{Tsirelson bound} \cite{tsirelson}. Going beyond the bipartite scenario, spatial quantum correlations have been analysed for multipartite scenarios as well \cite{bell_multi}. Other measures of spatial quantum correlations that have been studied  extensively are entanglement \cite{Schrodinger,horodecki_ent}, discord \cite{zurek,vedral}, and measurement-induced disturbances \cite{mid}.

Although temporal correlations too can provide a powerful quantum resource, their possible benchmarking has been studied much less. For example, Ref.~\cite{temporal_steer} shows that temporal steering allows us to identify the influence of a quantum measurement that has been performed in the past on future measurement outcomes. This type of influence of quantum measurements is thought to be the power behind certain quantum cryptographic protocols, including BB84 \cite{BB84}. Temporal quantum correlations also underlie a refined understanding of causality at the quantum level \cite{quantum_causality}. In addition, from the perspective of fundamental physics, it is important to understand how temporal correlations can be used to distinguish alternative physical theories of reality \cite{LG}. One question that naturally arises in this context is, can we quantify temporal quantum correlations in the same way as we quantify entanglement and can we set boundaries using CHSH-like inequalities?

Suppose two parties, Alice and Bob, conduct temporally ordered measurements on a single quantum system in a laboratory, as illustrated in Fig.~\ref{figure1}. Their individual measurement timings are non-intersecting, which means that Alice leaves the lab before Bob arrives. During her shift, Alice measures one of the two dichotomic observables, $A_1$ or $A_2$, and likewise, Bob measures one of the two dichotomic observables, $B_1$ or $B_2$. Now let us make a crucial assumption, namely that Alice can only conduct the above two measurements ($A_1$ and $A_2$) and that she cannot disturb the system and its natural dynamics in any other way. In this case, we can use joint probability distributions to describe the outcomes of the temporally separated measurements of Alice and Bob to differentiate between classical and quantum temporal correlations. In other words, we can ask, is there a quantity $S$ that can be used to distinguish between quantum and classical finite state generators.

The most celebrated work on temporal quantum correlations is perhaps the one done by Leggett and Garg \cite{LG,L}. Their inequalities can be violated quantum mechanically and can be used to characterise probabilistic hidden-variable models. More concretely, they considered two-time correlators between subsequent measurements of three $\pm 1$-valued observables. In contrast to space-like separated situations, it is not necessary to have more than one observable for each party. This condition implies that the observables used by Alice and Bob at different points in time do not need to commute in order to lead to quantum phenomena. Further studies on the characterisation of temporal quantum correlations, which consider a similar setup as that of Leggett and Garg, can be found in Refs.~\cite{guhne,Fritz,Hoffmann,multi_LG,Garg2025}.  Nevertheless, the success of previous characterisations of temporal quantum correlations has been limited.

As this paper illustrates, our difficulties with the benchmarking of temporal quantum correlations might be due to the fact that quantum and classical finite state generators are fundamentally different; both have their respective merits. On one hand, the output sequences of classical stochastic finite state generators can demonstrate a high level of complexity. On the other hand, classical measurements leave the classical system in a well-defined state, thereby essentially erasing all information about its history. This is not the case for quantum finite state generators, since generalised quantum measurements do not completely erase the quantum state of the measured system. Some information about the history of the quantum stochastic generator is maintained and can influence its future dynamics. As a result, classical short-term correlations can be more complex than the temporal correlations generated by quantum processes with comparable resources. To observe a significant difference between quantum and classical finite state generators, one needs to consider for example, temporal long-range correlations, like probabilities for words with multiple letters
\cite{AlRabsi2024}.

In addition, let us point out that temporal quantum correlations have been studied from many different perspectives in a variety of research areas. An example is the scrambling of information in condensed matter systems \cite{IS1,IS2}, quantum chaos \cite{qchaos,chaos}, and blackhole information \cite{bhole,susskind}, to name a few. Using Out-Of-Time-Ordered Correlators (OTOC) \cite{otoc,otoc_qm}, it has been shown that information can spread from one time-point to another through stochastic processes. This type of information scrambling is closely related to quantum chaos \cite{qchaos}, and the  mutual information can be bounded from below by the time-dependent change of OTOC \cite{mi_otoc}. Apart from the study of temporal correlations in terms of the evolution of quantum Heisenberg operators using OTOC in condensed matter and quantum chaotic systems, these correlations are also studied in systems following Markovian dynamics \cite{OTOC_markov}. 

In this paper, we consider CHSH scenarios with two subsequent measurements for each observer. It is shown that, in this case, the classical and quantum limits set by CHSH-like inequalities for temporal correlations are different from the limits obtained for spatial correlations. More concretely, we find that the CHSH parameter $S$ for temporal correlations of both classical and quantum stochastic generators can exceed the Tsirelson bound. From this we conclude that CHSH-like inequalities may not be the best candidate for benchmarking quantum temporal correlations. To show that this is indeed the case, we first review the general frameworks for the modeling of classical and quantum temporal stochastic processes in terms of Hidden Markov Models \cite{Rabiner1989,Xue2006,Vanluyten2008} and Hidden Quantum Markov Models \cite{Monras2011,Clark2015,AlRabsi2024}, respectively. Later, we compare the CHSH scores generated by these processes. Eventually, our studies illustrate that more complex measures than CHSH-like inequalities are required to identify the presence of non-classical temporal correlations. We show this by considering instead some longer range temporal correlations, rather than the immediate ones considered in the standard CHSH setting.  Here, we indeed find that the quantum processes have a noticable advantage over their classical counterparts.

The rest of the article is arranged as follows. Section \ref{sec2} briefly reviews the Bell-CHSH-like inequalities for temporal correlations in bipartite scenarios, which have already been considered by other authors. Sections \ref{sec3} and \ref{sec4} introduce general frameworks for classical and quantum stochastic finite state generators based on a single bit and a single qubit, respectively, building on previous work on Hidden Markov and Hidden Quantum Markov Models. In Section \ref{sec5}, we systematically search for possible violations of CHSH-like inequalities of these machines. Finally, in Section \ref{sec6}, we conclude our results.  

\section{Bell-CHSH inequalities} \label{sec2}

In this section, we review previous work regarding inequalities that have been proposed to distinguish quantum from classical for both spatial and temporal correlations. First we have a closer look at spatial quantum correlations.

\subsection{Spatial CHSH inequality} 

Suppose two spatially separated observers, Alice and Bob, share two quantum systems prepared in a joint quantum state $|\psi_{\rm AB} \rangle$. Each observers randomly performs one of two measurements, which we label $n,m \in \{1, 2\}$, on their respective subsystems.  Moreover, let us assume that each measurement has two possible outcomes, which we label $i,j = \pm 1$. In the following, we denote the corresponding observables for Alice's and Bob's measurements by $A_n$ and $B_m$, respectively. Then the expectation value of the product of their measurement outcomes equals
\begin{eqnarray} \label{E3}
	\langle A_n B_m \rangle &=& \sum_{i,j = \pm 1} i j \, p(ij|A_n B_m) \, ,
\end{eqnarray} 
where $p(ij|nm)$ is the probability of getting the outcomes $i$ and $j$ conditional on Alice and Bob measuring the observables $A_n$ and $B_m$.

In addition, let us assume that the observers consider the expression  
\begin{eqnarray} \label{E4}
	S &=& \left| \langle A_1 B_1 \rangle +  \langle A_1 B_2 \rangle + \langle A_2 B_1 \rangle - \langle A_2 B_2 \rangle \right|
\end{eqnarray} 
which is a function of the probabilities $ p(ij \lvert nm)$ and the shared state $|\psi_{\rm AB} \rangle$. If the results of the observables $n$ and $m$ are predefined, in other words, if the probabilities satisfy a local decomposition and the state $|\psi_{\rm AB} \rangle$ does not contain any entanglement \cite{horodecki_ent}, then necessarily 
\begin{eqnarray} \label{E4b}
	S &\leq & 2 \, .
\end{eqnarray}
The above equation is the celebrated CHSH inequality \cite{chsh}. Any test of this inequality assumes that the two local observers have the free-will to choose between two alternative projective measurements corresponding to the observables they have chosen. The possibility of violating of the above CHSH inequality {\em without} local observers who choose between alternative projective measurements was reported in Ref.~\cite{bell_povm}, thereby signifying that a suitable generalised measurement is equivalent to a random selection between two alternative projective measurements.

Let us assume that Alice and Bob measure observables $A_n$ and $B_m$ which have been randomly chosen from the sets $\{A_1,A_2\}$ and $\{B_1,B_2\}$, respectively. As these observables are quantum mechanical in nature, the expectation values of their products equal
\begin{eqnarray} \label{E4c}
\langle A_nB_m \rangle &=& \langle \psi_{\rm AB} | A_n B_m |\psi_{\rm AB} \rangle \, .
\end{eqnarray}
Using this equation, one can show that the value of $S$ can go beyond the classical limit of 2 for quantum mechanical systems. For a maximally entangled state shared between Alice and Bob, the value of $S$ can be as high as $2\sqrt{2}$, both theoretically and experimentally \cite{chsh, Clauser_1978}. Overall, it was found that  
\begin{eqnarray} \label{Eq:CHSH_quantum}
S & \leq& 2 \sqrt{2} 
\end{eqnarray}
for all possible quantum states $|\psi_{\rm AB} \rangle$. A violation of the Bell-CHSH inequality in Eq.~(\ref{E4b}) by the shared state between Alice and Bob signifies the presence of non-classical correlations between them. In other words, the CHSH factor $S$ is an entanglement witness.

A maximum violation of the CHSH inequality in Eq.~(\ref{E4b}) is obtained, when the observers measure spin components in the $x$-$z$ plane of the Bloch sphere and Alice and Bob share a maximally entangled state. Suppose $|i_{\rm X} \rangle$ with $i=0,1$ and ${\rm X}={\rm A},{\rm B}$ are basis states of single qubits and $|\psi_{\rm AB} \rangle = |\phi^+\rangle $ with
\begin{eqnarray} \label{E2}
|\phi ^{+}\rangle &=& (|0_{\rm A} 0_{\rm B} \rangle + |1_{\rm A} 1_{\rm B} \rangle) /\sqrt{2} \, .
\end{eqnarray}
Moreover, we assume in the following that  
\begin{eqnarray}
A_n &=& \cos \theta_{\rm A}^{(n)} \, \sigma_z + \sin \theta_{\rm A}^{(n)} \, \sigma_x \, , \notag \\
B_m &=& \cos \theta_{\rm B}^{(m)} \, \sigma_z + \sin \theta_{\rm B}^{(m)} \, \sigma_x
\end{eqnarray}
with the angles $\theta_1^{(\rm A)} = 0$ and $\theta_2^{(\rm A)} = \pi/2$ for Alice and $ \theta_1^{(\rm B)} = -\pi/4$ and $\theta_2^{(\rm B)} = \pi/4$ for Bob. In this case, it can be shown that the calculation of the expectation values of the product of the above observables using Eq.~(\ref{E4c}) yields $S = 2\sqrt{2}$. Bell inequalities involving only the terms in Eq.~(\ref{E4c}), such as the CHSH inequality, are called correlation inequalities. Therefore, a violation of the CHSH inequality also signifies the presence of measurement correlations that can only be obtained from an entangled state; they can never be produced by a separable state. Bell-CHSH inequalities have also been studied for multipartite scenarios beyond
two-party settings \cite{bell_multi}. 

\subsection{Temporal CHSH inequality}

The key difference in the case of temporal correlations is that Alice and Bob act on the same qubit, rather than acting on their own and comparing measurement results afterwards.  In particular, we assume that Alice and Bob both perform individual measurements on the qubit which has initially been prepared in a state $| \psi \rangle$. Naturally, this creates a subtlety \textemdash namely the temporal ordering of Alice's and Bob's measurements becomes important. This is not the case when they measure on two spatially separated qubits. To account for this, we replace the CHSH parameter $S$ in Eq.~(\ref{E4}) in the following by 
\begin{align} \label{Eq:S_corr}
	S = | C_{11} + C_{12} + C_{21} - C_{22} | \, .
\end{align}
The {\em correlators} $C_{nm}$ in this equation can be written as
\begin{align} \label{Eq:Corr_def}
	C_{nm} = & \frac{1}{2} \langle \{A_n, B_m\} \rangle 
\end{align}
with $\{A_n, B_m\} = A_n B_m + B_m A_n$. This expression can be calculated using Eq.~(\ref{E3}), if the operators $A_n$ and $B_m$ are projectors. 
The expression in Eq.~(\ref{Eq:S_corr}) effectively averages over the two possible temporal orderings. Note also that the above definition of $S$ reverts back to the standard definition of $S$ in Eq.~(\ref{E4}) when $A_n$ and $B_m$ commute with each other, i.e.~when the temporal ordering of the measurements performed by Alice and Bob does not need to be taken into account.

Temporal correlations have already been studied in a variety of ways. For example, Ref.~\cite{Hoffmann} 
considers projective measurements on a single quantum system in a temporal version of the CHSH scenario and infers fundamental limitations of temporal quantum correlations. By deriving the validity of the Tsirelson bound and the impossibility of Popescu--Rohrlich-box behaviour, the authors of Ref.~\cite{PR_box} show that a set of correlators can appear in the temporal CHSH scenario if and only if it appears in the conventional CHSH setup. Let us consider a single quantum system with a Hilbert space $\mathcal{H}$ and a Hamiltonian $H$. On this system, the two parties Alice and Bob measure two dichotomic  Hermitian observables $A$ and $B$ with values $\pm 1$ at times $t_a$ and $t_b$, respectively. If the observables have the property $A^2 = B^2 = \mathds{1}$, then one can show using the Born rule that the joint probabilities for Alice's measurement outcomes to be $r\in \{-1, 1\}$ and for Bob's measurements outcomes to be $s\in \{-1, 1\}$ can be written as
\begin{eqnarray}
	P(r, s) &=& \biggl <\psi \bigg | \frac{\mathds{1} + rA}{2}. \frac{\mathds{1} + sB}{2}. \frac{\mathds{1} + rA}{2} \bigg | \psi \bigg > \nonumber\\
	&=& \frac{1}{4} + \frac{1}{4}r\langle\psi |A| \psi\rangle + \frac{1}{8}s\langle\psi |B| \psi\rangle \nonumber\\
	&&+ \frac{1}{8}rs\langle\psi |\{A, B\}| \psi\rangle + \frac{1}{8}s\langle\psi |ABA| \psi\rangle \, . ~~
\end{eqnarray}
Here $|\psi\rangle$ is the initial state of the quantum system and $(\mathds{1} + rA)/2$ and $(\mathds{1} + sB)/2$ are the projection operators onto the $\pm$-eigenspaces for Alice and Bob, respectively. Having derived the joint probabilities, the correlators can be expressed as
\begin{eqnarray} \label{Eq:Correlator}
	C &\equiv& \sum_{r,s} rs~P (r, s)\nonumber\\
	&=& P (+1, +1) + P (-1, -1) -  P (-1, +1) - P (+1, -1)\nonumber\\
	&=& \frac{1}{2}\langle \psi |\{A, B\}| \psi\rangle \, .
\end{eqnarray}
Once the stage is setup for sequential measurements on a single qubit, we can derive the temporal CHSH inequality in a manner analogous to the spatial case, using again the expression in Eq.~(\ref{Eq:S_corr}) to calculate $S$ but with $C_{nm}$ now refering to the correlator between measurement outcomes of Alice's and Bob's observables $A_n$ and $B_m$, respectively. Using some novel techniques, the authors in \cite{Hoffmann} have shown that for a single qubit the maximal achievable value of $S$ in the temporal CHSH scenario is the Tsirelson bound of $2\sqrt2$, if Alice and Bob can freely select their observables from the set $\{A_1, A_2\}$ and from $\{B_1, B_2\}$, respectively.

\subsection{Other measures of temporal correlations}

So far, we briefly described how one can obtain a temporal inequality in the CHSH scenario. However, quantum temporal correlations can also be studied using a more general approach, as shown in Ref.~\cite{guhne}. Considering a single quantum system and performing sequential measurements on it can yield temporal quantum correlations that are not achievable for macro-realistic and non-contextual models. This general approach helps us with the full characterisation of temporal correlations in the simplest Leggett-Garg scenario and in the sequential measurement scenario associated with the most fundamental proof of the Kochen-Specker theorem \cite{kochen}. Suppose $\langle A_iA_j\rangle$ denotes the expectation value of the product of the observables $A_i$ and $A_j$ when the observable $A_i$ is measured first and $A_j$ later. Using the product terms mentioned above, one can define a correlation function $S_5$ such that 
\begin{eqnarray}
	S_5 &=& \langle A_1A_2\rangle_{\rm seq} + \langle A_2A_3\rangle_{\rm seq} + \langle A_3A_4\rangle_{\rm seq}\nonumber\\
	&&+ \langle A_4A_5\rangle_{\rm seq} - \langle A_5A_1\rangle_{\rm seq} \, ,
\end{eqnarray}
and find classical and quantum bounds on temporal quantum correlations.
Evaluating the above mentioned quantity, it can be shown that the inequality $S_5 \leq 3$ holds for macro-realistic and non-contextual models. 
To find the quantum bound for $S_5$, one can express the expectation value of the product of the observables as
\begin{eqnarray}
	\langle A_iA_j\rangle_{\rm seq} &=& \frac{1}{2}[\mbox{Tr}(\rho A_iA_j) + \mbox{Tr}(\rho A_jA_i)] \nonumber\\
	&=& \frac{1}{2}[\mbox{Tr} (\rho \{A_i, A_j \})] \, .
\end{eqnarray}
Using this expression and doing optimisation leads to 
\begin{eqnarray}
	S_5 &\leq & \frac{5}{4} \left(1 + \sqrt{5} \right) \approx 4.04 \, ,
\end{eqnarray} 
with the right hand side of this equation signifying the maximum quantum value of $S_5$ \cite{guhne}.
However, if the observables in each measurement sequence commute, then the quantum bound of $S_5$ is known to be approximately $3.94$. Eqs.~(\ref{Eq:S_corr}) and (\ref{Eq:Correlator}) show that, unlike a spatial correlator which is given by the expectation value of the tensor products of the observables, the temporal correlator used for exploring quantum temporal correlations is given by half the expectation value of the anti-commutator of the observables. Although there are many different possible definitions of the CHSH score $S$, we choose to focus in the following on the one given in Eq.~(\ref{E4}) which is enough to demonstrate that comparing quantum and classical temporal correlations is not straightforward due to the relatively high complexity of classical stochastic finite state generators.

\section{A general framework for the modelling of classical temporal stochastic processes} \label{sec3}

Next, let us have a closer look at classical probabilistic finite state generators with two possible output symbols and two possible internal states. For pedagogical reasons, we distinguish two types of machines. The first one are so-called {\it Markov models}. These randomly generate an output symbol while the machine transitions into the corresponding internal state. In contrast to this, for {\it Hidden Markov Models}, the random output symbol is no indication of the internal state that the machine transitions into. In case of a read-out of a symbol on a Hidden Markov Model, the internal state of the machine remains hidden. In the remainder of this section, we introduce the mathematical tools needed to describe these two types of classical stochastic generators before studying their temporal correlations in Section \ref{sec5}.

\subsection{Markov Models}

Markov models are a specific class of probabilistic finite state machines that do not have an internal memory of their state history. The evolution of their internal state only depends on the current internal state and a set of fixed transition probabilities. The  processes generated by such machines are therefore also known as Markov processes. In the following, we consider classical one-bit machines which can only be prepared in one of two possible internal states $x = \pm 1$. In addition, we denote the possible outputs of the machine by $i=\pm 1$. Now suppose that we are only interested in the expectation values and the out-of-time-order correlators of subsequent readouts of the machine. In this case, we do not need to keep track of all possible trajectories that the stochastic generator can produce. We only need to keep track of probabilities to find the machine after a certain number of steps in a certain internal state $x$.

In the following, we describe the current state of a machine whose history is not known by a two-dimensional vector $\eta$ of the form
\begin{eqnarray} \label{Etrans}
	\eta &=& (p_{-1}(\eta),p_{+1}(\eta))^{\rm T}
\end{eqnarray}
where $p_x(\eta)$ with $p_{-1}(\eta) + p_{+1}(\eta) = 1$ denotes the probability to find a Markov Model prepared in $\eta$ in $x$.
In addition, we characterise concrete Markov Models by transition matrices $T$ of the form 
\begin{eqnarray} \label{Etrans22}
	T &= & \left( \begin{array}{cc} p(-1|-1) & p(-1|+1) \\ p(+1|-1) & p(+1|+1) \end{array} \right) .
\end{eqnarray}
Here $p(y|x)$ is the conditional probability of the machine transitioning from the state $x$ into the state $y$. Using this notation, we find that the state of the machine changes from $\eta$ into $T \eta$,
\begin{eqnarray} \label{Etrans22c}
	\eta & \longrightarrow & T\eta \, ,
\end{eqnarray}
 during a single read out or time step. The above probabilities $p(y|x)$ must be properly normalised to always yield a normalised state, i.e.~we require that
\begin{eqnarray} \label{Etrans22b}
	\sum_{y=\pm 1} p(y|x) &=& 1
\end{eqnarray}
for $x=-1$ and for $x=+1$. Since each transition matrix depends on four conditional probabilities which are constrained by the two conditions in Eq.~(\ref{Etrans22b}), $T$ can be fully characterised by two independent real parameters $a,b \in [0,1]$. More concretely, we find that $T$ can always be written as 
\begin{eqnarray} \label{Etrans4}
	T &= & \left( \begin{array}{cc} a & 1-b \\ 1-a & b \end{array} \right) \, ,
\end{eqnarray}
with $a = p(-1|-1)$ and $b = p(+1|+1)$. To numerically sample a large set of all possible Markov models, all we need to do is to generate a large number of random parameters $a$ and $b$ and compare the output sequences of the corresponding machines.

To calculate conditional probabilities and the out-of-time-order correlators of subsequent measurements, it is sometimes convenient to decompose the transition matrix $T$ into two transition matrices $T^{(\pm 1)}$ with 
\begin{eqnarray} \label{Etrans9}
	T^{(-1)} = \left( \begin{array}{cc} a & 1-b \\ 0 & 0 \end{array} \right) , ~~ 
	T^{(+1)} = \left( \begin{array}{cc} 0 & 0 \\ 1-a & b \end{array} \right) \, .
\end{eqnarray}
These two transition matrices can be used to calculate the state of the one-bit machine conditional on obtaining measurement outcome $i$ after the machine has initially been prepared in $\eta$. More concretely, the machine changes such that
\begin{eqnarray} \label{Etrans22d}
	\eta & \longrightarrow & {T^{(i)} \eta \over P (T^{(i)} \eta)} \, ,
\end{eqnarray}
where $P(T^{(i)} \eta)$ with
\begin{eqnarray} \label{Etrans22dd}
P(T^{(i)} \eta) &=& (1,1)T^{(i)} \eta 
\end{eqnarray}
tells us the probability of the machine evolving with $T^{(i)}$. Given the construction of the above transition matrices, one can easily check that 
\begin{eqnarray} \label{Etrans8}
T &=& T^{(-1)} + T^{(+1)} 
\end{eqnarray}
and that, for Markov Models, the state of the machine equals $x=i$, if the measurement outcome $i$ is obtained. 

\subsection{Hidden Markov Models}

Next we turn our attention to Hidden Markov Models, which are a generalisation of Markov Models. The main difference is that the output symbol $i$ is no longer an indication of the internal state $x$ of the machine. In other words, the internal state of the machine, which determines the probability for obtaining a certain output symbol and for changing the internal state of the machine in the next read out, remains hidden. Hidden Markov Models with one internal bit are characterised by a set of eight parameters $p_i(y|x)$ where $i$, $x$ and $y$ can each assume two different values. Without restrictions, we assume in the following that these values are $-1$ and $+1$. Moreover, we identify $p_i(y|x)$ with the conditional probability of obtaining the output $i$ and transitioning into the state $y$ in the next time step given the internal state of the machine equals $x$. 

As we did with the Markov Models in the previous subsection, we characterise concrete Hidden Markov Models in the following by transition matrices $T^{(\pm 1)}$ of the form 
\begin{eqnarray} \label{Etrans22x}
	T^{(i)} &= & \left( \begin{array}{cc} p_i(-1|-1) & p_i(-1|+1) \\ p_i(+1|-1) & p_i(+1|+1) \end{array} \right) \, .
\end{eqnarray}
Probability theory tells us that the conditional probabilities $p_i(y|x)$ must obey the completeness relation
\begin{eqnarray} \label{Etrans22xx}
\sum_{y=\pm 1} \sum_{i=\pm 1} p_i(y|x) &=& 1
\end{eqnarray}
for $x = -1$ and for $x=+1$.  This relation ensures that a machine prepared in $x$ always generates an output symbol $i$ and always transitions into one of the two possible states $y = \pm 1$. In Section \ref{sec4}, we will use the above transition matrices to calculate for example transition probabilities and the expectation values for certain output sequences.

Since the internal state of the machine remains unknown even when we keep track of its output symbols, we describe the internal state of Hidden Markov Models in the following by the vector $\eta$ in Eq.~(\ref{Etrans}) with $p_x(\eta)$ denoting again the probability of $x$ currently being the internal state of the machine. If the output symbol $i$ of the machine is ignored, then this state changes from $\eta$ to $T\eta$, as in Eq.~(\ref{Etrans22c}), if we define the total transition matrix $T$ as in Eq.~(\ref{Etrans8}) as the sum of the transition matrices $T^{(0)}$ and $T^{(1)}$. Because of Eq.~(\ref{Etrans22xx}), one can show that $(1,1) T\eta  = 1$ which means that the resulting state is normalised, as should be the case. However, if we know that a certain output symbol $i$ has been obtained, then the state of the Hidden Markov Model changes according to Eq.~(\ref{Etrans22d}). By construction, the probability of obtaining the outcome $i$ given a machine prepared in $\eta$ still equals $P (T^{(i)} \eta)$ in Eq.~(\ref{Etrans22dd}). 

Nevertheless, as we shall see in the remainder of this paper, the temporal correlations that Hidden Markov Models can produce can be more complex than the correlations of  Markov Models. This is due to the presence of additional pathways for the machine to evolve internally while generating a sequence of seemingly random but correlated output symbols. Moreover we notice that the transition matrices $T^{(-1)}$ and $T^{(+1)}$ of Hidden Markov Models are characterised by eight conditional probabilities $p_i(y|x) $ with two constraints (cf.~Eq.~(\ref{Etrans22xx})). Hence their parametrisation requires six independent parameters $a$, $b$, $c$, $d$, $e$ and $f$. Taking this into account, we write $T^{(-1)}$ and $T^{(+1)}$ in the following as
\begin{eqnarray} \label{Etrans99}
	T^{(-1)} &=& \left( \begin{array}{cc} a & d \\ b & e \end{array} \right) \, , \notag \\
	T^{(+1)} &=& \left( \begin{array}{cc} c & f \\ 1-a-b-c & 1-d- e-f \end{array} \right) 
\end{eqnarray}
with the range of the parameters $a$ to $f$ given by
\begin{eqnarray} \label{Etrans100}
&& a,d \in (0,1) \, , ~~ b \in (0,1-a) \, , ~~ c \in (0,1-a-b) \, , \notag \\
&& e \in (0,1-d)\, , ~~ f \in (0,1-d-e) \, . 
\end{eqnarray}
It is relatively straightforward to check that the conditional probabilities $p_i(y|x)$ shown in Eq.~(\ref{Etrans99}) obey the constraint in Eq.~(\ref{Etrans22xx}) for $x=- 1$ and for $x=+1$. 

\section{A general framework for the modelling of quantum temporal stochastic processes} \label{sec4}

If we replace the one-bit machines used to implement Hidden Markov Models by quantum machines containing a single qubit and analyse their output sequences in the case of sequential measurements, we obtain so-called Hidden Quantum Markov Models \cite{Monras2011,Clark2015,AlRabsi2024}. These are the quantum analogues of the Hidden Markov Models that we introduced in the previous section. As we illustrate in Section \ref{sec5} of this paper, their output sequences are qualitatively different. In this section, we parametrise the Kraus operators \cite{q_channel, Nielsen:2012yss} which can be used to characterise all possible Hidden Quantum Markov Models. Each Kraus operator describes a generalised measurement that is performed on a single qubit and which reveals a random output symbol $i = \pm 1$ and causes the internal state of the machine to evolve. As we shall see below, Hidden Quantum Markov Models are characterised by seven independent real parameters, which is more than the number of parameters required to characterise a Hidden Markov Model. This observation hints at an increase in complexity over the classical machines which have at most six independent real parameters.

\subsection{Kraus operators} \label{secK}

Suppose a single qubit is initially prepared in a quantum state $|\psi \rangle$ of the form
\begin{eqnarray} \label{E9}
|\psi \rangle &=& \alpha \, |-1 \rangle + \beta \, |+1 \rangle \, .
\end{eqnarray}
Here $\alpha$ and $\beta$ are complex coefficients with $|\alpha|^2 + |\beta|^2 = 1$ and the states $|-1 \rangle$ and $|+1 \rangle$ are two orthogonal basis states with $\langle -1|+1 \rangle = 0$ which replace the classical internal states $x=\pm 1$ of Hidden Markov Models. Performing a generalised measurement on the quantum state $|\psi \rangle$ leads to two possible outcomes. Since we are later interested in the CHSH inequality, we denote these again by $i=\pm 1$ and assign a Kraus operator $K^{(i)}$ to each $i$. For example, if the outcome of a measurement on $|\psi \rangle$ equals $i$, then the internal quantum state changes such that
\begin{eqnarray}
|\psi \rangle &\longrightarrow & {K^{(i)} |\psi \rangle \over P(K^{(i)} \psi)} \, .
\end{eqnarray}
This equation has many similarities with Eq.~(\ref{Etrans22d}). The main difference is that the state vector $\eta$ and the transition matrix $T^{(i)}$ have now been replaced by the state vector $|\psi \rangle$ and the Kraus operator $K^{(i)}$. Moreover, the probability to obtain the outcome $i$ now equals
\begin{eqnarray} \label{E29}
P(K^{(i)} \psi) &=& \| K^{(i)} |\psi \rangle \|^2 \, ,
\end{eqnarray}
which replaces Eq.~(\ref{Etrans22dd}) for Markov and Hidden Markov Models. This acts as the observable for a single measurement in the HQMM, emitting the output symbol $i$. 

In the remainder of this section, we parametrise all possible sets of Kraus operators $K^{(i)}$ and hence also all possible observables $A$ for generalised measurements on a single qubit. From quantum physics, we know that the only condition the Kraus operators $K^{(i)}$ have to obey is the completeness relation
\begin{eqnarray} \label{C1}
\sum_{i=\pm 1} K^{(i) \dagger} K^{(i)} &=& I
\end{eqnarray}
where $I = \sum_{x =\pm 1} |x \rangle \langle x|$ denotes the identity operator. The most general Kraus decomposition for an (unobserved) quantum channel results in a set of four operators, with 12 free parameters.  However, as we are strictly concerned with an observed readout with two possible output symbols, we find that there are a total of 8 free parameters, which can further be reduced to 7 through accounting for a global phase ambiguity with no physical consequences \cite{Nielsen:2012yss}. This is found by enforcing Eq.~(\ref{C1}) for a basis of two Kraus operators.

Physically performing a generalised measurement on a qubit requires bringing it in contact with an additional auxiliary qubit. This second qubit is part of the measurement apparatus and does not contain any information so that the evolution of the Hidden Quantum Markov Model during a read out step depends purely on the current internal state $|\psi \rangle$ of the machine. For simplicity, we assume in the following that the initial state of the auxiliary qubit always equals $|-1 \rangle$. Next, a unitary operation $U$ should be performed which evolves the state $|-1,-1 \rangle$ into $|a \rangle$ and the state $|+1,-1 \rangle$ into $|b \rangle$. As a result,
\begin{eqnarray} \label{E32}
|\psi \rangle &\longrightarrow & \alpha |a \rangle + \beta |b \rangle \, ,
\end{eqnarray}
thereby entangling the internal state of the Hidden Quantum Markov Model with the auxiliary qubit. This transformation can be physically realised as long as 
\begin{eqnarray} \label{E21b}
\langle a|a \rangle = \langle b|b \rangle =1 \, , ~~
\langle a|b \rangle &=& 0 \, .
\end{eqnarray}
The above completeness and orthogonality relations are required, since two distinguishable states, like the state $|-1,-1 \rangle$ and state $|+1,-1 \rangle$, necessarily evolve into two distinguishable normalised states when experiencing the same unitary evolution.  

To complete the generalised measurement on the memory qubit of the Hidden Quantum Markov Model and to obtain one of the two possible measurement outcomes $i=\pm 1$, we finally perform a projective measurement on the ancilla qubit. As the following two subsections illustrate, this measurement can reveal some information about the state $|\psi \rangle$ of the memory qubit. How much information is obtained can be controlled by carefully choosing the basis for the projective measurement. In general, the internal state of the machine remains in a superposition of the states $|- 1 \rangle$ and $|+ 1 \rangle$ with coefficients which depend on the coefficients of the initial state $|\psi \rangle$. In this way, information is carried forward. This is different from Markov and Hidden Markov Models where a read out necessarily erases all information about the history of the machine. Notice also, that one could let the memory qubit interact with a collection of ancillas when implementing a generalised measurement, but as shown by the Stinespring theorem \cite{stinespring}, this does not increase the complexity of the resulting dynamics.

\subsection{Projective measurements} \label{sec3A}

Suppose the transition in Eq.~(\ref{E32}) is the result of a so-called double encoding, which is a crucial step when carrying out the quantum teleportation of a single qubit or other quantum information tasks. In this case, $|a \rangle = |-1,-1\rangle$, $|b \rangle = |+1,+1 \rangle$ and 
\begin{eqnarray} \label{E32z}
|\psi \rangle & \longrightarrow & \alpha  |-1,-1\rangle + \beta  |+1,+1\rangle \, .
\end{eqnarray}
Next we perform a projective measurement on the ancilla qubit with the projectors $P^{(i)}$ given by
\begin{eqnarray} \label{E12}
P^{(i)} &=& |i \rangle \langle i| \, .
\end{eqnarray} 
Here $i$ corresponds to the respective measurements outcome $i=-1$ or $i=+1$, respectively. Given the above scenario, the final state of the memory qubit of the Hidden Quantum Markov Model equals $|-1 \rangle$ with probability $|\alpha|^2$ and $|+1 \rangle$ with probability $|\beta|^2$. Overall, the initial qubit state has been erased and a projective measurement in the basis of the $|\pm 1 \rangle$ states has been performed on the memory qubit. Hence $K^{(i)} = P^{(i)}$ with the projectors $P^{(i)}$ given in Eq.~(\ref{E12}). Proceeding analogously, while replacing the states $|\pm 1 \rangle$ by other basis states, other projective measurements can be realised on the memory qubit.

\subsection{Non-projective measurements} \label{sec3C}

Next we have a closer look at generalised measurements on the qubit, of which the projective measurements are a special example. First we notice that it does not matter in which basis we measure the auxiliary qubit in, since any change of basis can be compensated for by adjusting the states $|a \rangle$ and $|b \rangle$ accordingly. We therefore allow a general operation on our combined system giving us a state of the form of Eq.~(\ref{E32}).  We decompose the states $\ket{a}$ and $\ket{b}$ as
\begin{eqnarray}
    \ket{a} &= & \ket{a_{-1}}_\mathrm{int} \ket{-1}_\mathrm{anc} + \ket{a_{+1}}_\mathrm{int} \ket{+1}_\mathrm{anc} \, ,
\end{eqnarray}
where $\ket{a_i}_\mathrm{int} = \prescript{}{\mathrm{anc}}{\braket{i | a}}$.  This allows us to rewrite the state from Eq.~(\ref{E32}) as
\begin{eqnarray}
    \ket{\psi} &= & \sum\limits_{i=\pm 1} \left(\alpha \ket{a_1}_\mathrm{int} + \beta \ket{b_i}_\mathrm{int} \right) \ket{i}_\mathrm{anc}
\end{eqnarray}
Based on this, we may idenitfy the Kraus operators as
\begin{eqnarray}
    K^{(i)} & = & \ket{a_i}\bra{-1} + \ket{b_i}\bra{+1} \, .
\end{eqnarray}
We then may sample over states $\ket{a_i}$ and $\ket{b_i}$, enforcing satisfaction of the completeness relation and orthogonality of these states, to generate our family of Kraus operators.  In doing so, as mentioned above, we arrive at 8 free parameters, which can be reduced to 7 with the absorption of a global phase ambiguity.
 

\section{Comparing the temporal CHSH scores of classical and quantum stochastic processes} \label{sec5}

Now we can finally compute the CHSH score $S$ in Eq.~(\ref{Eq:S_corr}) for temporal processes and assess their limits for different machines. We begin by identifying the correlators necessary for their calculation, before sampling large numbers of output sequences.

\subsection{Constructing the correlators}

As mentioned before, in the following, we consider correlators to describe the properties of temporal processes. For the classical machines, we can simply use the transition operators to construct the observables, substituting the probabilities of each event into Eq.~(\ref{Etrans22x}). For the non-projective quantum cases though, we cannot use the discrete observables discussed in Eq.~(\ref{Eq:Corr_def}).  The reason is that the temporal ordering does not cancel identically in general for a Kraus evolution where the order of measurement between Alice and Bob is swapped, whereas in the projective case it does.  We therefore construct a joint observable as
\begin{eqnarray} \label{Eq:Kraus_obs}
    \langle O(A_n,B_m) \rangle & = & \sum\limits_{i,j=\pm 1} ij \, ||B_m^{(j)} A_n^{(i)} \ket{\psi}||^2 \, , 
\end{eqnarray}
where we have defined the joint observable $O(A_n,B_m)$ as the sequence of events where Alice measures first with her Kraus operator in basis $n$ followed by Bob measuring with his Kraus operator in basis $m$.  In other words, we are only able to establish the joint observable between Alice and Bob's measurements, due to their intrinsic correlation.  In the case that the operators commute, the above expression on the right hand side of the equation trivially returns to the standard expectation value form.  We will thus work with Eq.~(\ref{Eq:Kraus_obs}) going forward to ensure we are correctly accounting for the correlators in our results.

\subsection{Calculating expectation values and correlators}

Now we have the methodology to sample many instances of each type of machine using Monte Carlo methods. Crucial in this process is determining the expectation values of certain outcomes using Eqs.~(\ref{E3}) and (\ref{Eq:Kraus_obs}), respectively, and substituting the resulting expressions into Eqs.~(\ref{Eq:S_corr}) and (\ref{Eq:Corr_def}). In our sampling process, we need to create the basis choices for Alice and Bob, before then calculating the probability of all outcomes. This means applying the specific transition/Kraus operators for a given basis choice and outcome to determine their likelihood and corresponding CHSH score. In line with Eq.~(\ref{E3}), we shall label the transition operators and Kraus operators of Alice $A$, while Bob's are labelled $B$.  Moreover, the subscript corresponds to their basis choice.  As such, we sample two basis options for Alice and Bob, $A_{1,2}$ and $B_{1,2}$, respectively, and then determine expectation values $\langle A_n B_m \rangle$ and $\langle B_m A_n \rangle$. 

More concretely, we proceed as follows. In each iteration of our sampling process, we generate two basis choices for Alice and Bob each, forming the sets of transition matrices or Kraus operators that they shall use to establish the CHSH score.  We then, utilising the formalisms described in the previous sections, determine each of the expectation values required to construct the correlators and thus find the CHSH score of the sampled sets of transition matrices or Kraus operators.  We repeat this many times, to construct a distribution of CHSH scores for many different parameterisations of machines used by Alice and Bob.  The result is that we approach the ultimate limits of the machines with a sufficiently large amount of sampling, but also determine the general behaviour, e.g.~how likely it is that the machines produce the highly correlated behaviour needed for large CHSH scores.  This gives some insight into the typical dynamics of the characterisation.

\subsection{Comparing the CHSH score between quantum and classical processes}

\begin{figure}[t] 
	\centering \includegraphics[width=\linewidth]{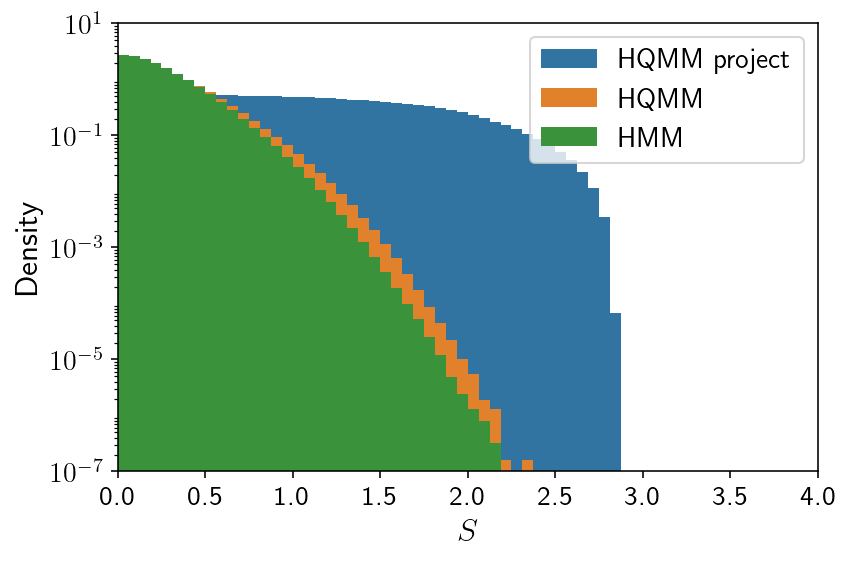}
	\caption{CHSH scores of 1-bit HMMs, 1-qubit HQMMs and 1-qubit HQMMs with projective-like Kraus operators, generated from sampling $10^8$ different machines.  We see that there is a moderate increase in relative complexity, characterised by the CHSH score, going from the classical to quantum machines, but it is rare to find any surpassing the typical limit of 2.  When we restrict the quantum machines to projective-like however, we see a more expected distribution, saturating the usual bound up to $2 \sqrt{2}$.
	}
	\label{Fig:CHSH} 
\end{figure}

With the sampling procedure highlighted in the previous subsection, we can build a distribution of CHSH scores to see how the different machines perform.  We summarise our findings in Fig.~\ref{Fig:CHSH}. Here, we plot the samples in a histogram, with the values being normalised by the bin width and total number of samples to provide a density.  In both the quantum and classical cases, we assume the system is initially prepared in the $-1$ state with unit probability, as with sufficient sampling the results should be initial state independent.  We see that the general temporal quantum evolutions rarely manages to surpass $S=2$, but can achieve this in principle, with the highest observed value being $S\sim2.31$.  For the projective case, we see that saturating the CHSH inequality with $S=2 \sqrt{2}$ is more likely than in the general temporal case, with values of $S$ very close to $2\sqrt{2}$ observed.   For the classical machine, we see generally similar behaviour to the general quantum temporal process, with the highest observed value of $S\sim2.12$.  This means that the classical temporal process is capable of surpassing $S=2$, unlike the spatial classical case.

We note here that the projective case of the HQMMs should be included in the general HQMMs.  Hence the bound achieved by these also applies equally to the HQMMs.  We separate them to illustrate the point that a lot of the HQMMs, due to their larger parameter space and less efficient sampling, are not good candidates for producing correlated outputs and do not take advantage of their full potential, but despite this if they are appropriately parameterised they can achieve high scores.  We plot the density of the results of our sampling to demonstrate these different rates.

\begin{figure*}[t] 
	\centering \includegraphics[width=0.32\linewidth]{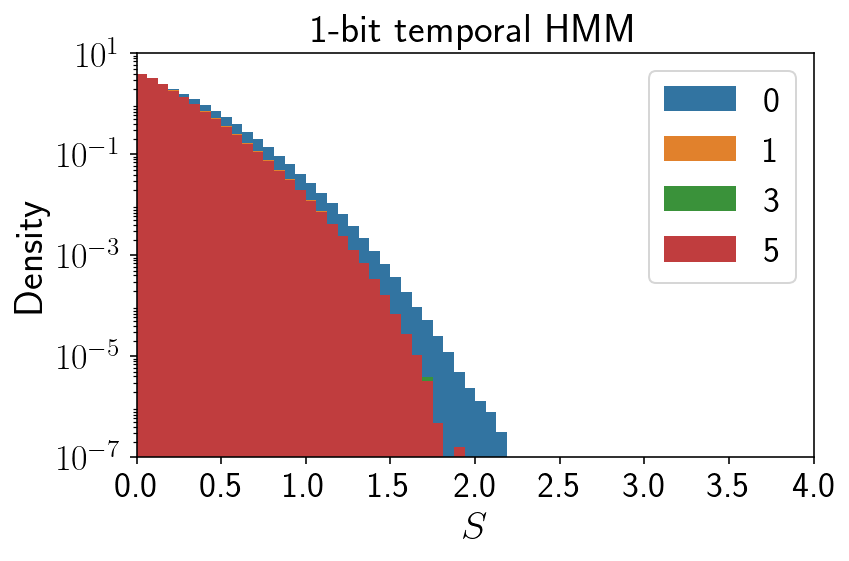} \includegraphics[width=0.32\linewidth]{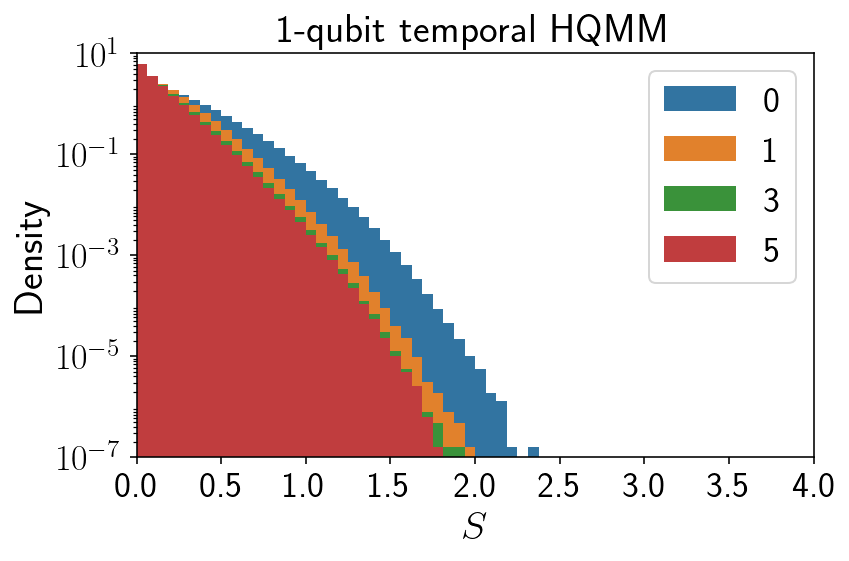}  \includegraphics[width=0.32\linewidth]{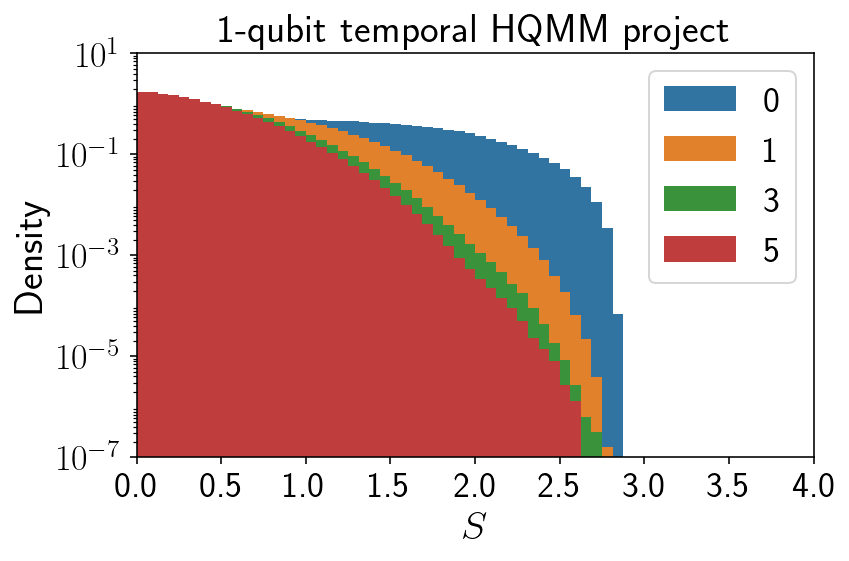}
	\caption{CHSH scores for the three classifications of machines, again sampled from $10^8$ different occurences.  We now allow for a random mapping between the operations of Alice and Bob by a third user, Charlie, who has his own set of transition matrices/Kraus operators.  This degrades the correlation between Alice and Bob's measurements.  We see that the quantum machines, particularly in the projective case, are better at maintaining this correlation, even under Charlie's action.
	}
	\label{Fig:CHSH_time_lag}
\end{figure*}

Perhaps more surprising is that the classical process can even surpass the limit of $S = 2 \sqrt{2}$, although only in very specific scenarios.  This can be seen through a simple example of a MC.  Let Alice's transition matrices be
\begin{eqnarray} \label{Eq:Example_MC_Alice}
	A_+^{(1)} &= & \left(\begin{array}{c c} 0 & 0 \\ 0 & 0 \end{array} \right) \, , \quad A_-^{(1)} = \left(\begin{array}{c c} 0 & 0 \\ 1 & 1 \end{array} \right) \, , \nonumber \\
	A_+^{(2)} &= & \left(\begin{array}{c c} 1 & 1 \\ 0 & 0 \end{array} \right) \, , \quad A_-^{(2)} = \left(\begin{array}{c c} 0 & 0 \\ 0 & 0 \end{array} \right) \, ,
\end{eqnarray}
while Bob takes
\begin{eqnarray} \label{Eq:Example_MC_Bob}
	B_+^{(1)} & = & \left(\begin{array}{c c} 0 & 0 \\ 0 & 0 \end{array} \right) \, , \quad B_-^{(1)} = \left(\begin{array}{c c} 0 & 0 \\ 1 & 1 \end{array} \right) \, , \nonumber \\
	B_+^{(2)} & = & \left(\begin{array}{c c} 1 & 0 \\ 0 & 0 \end{array} \right) \, , \quad B_-^{(2)} = \left(\begin{array}{c c} 0 & 0 \\ 0 & 1 \end{array} \right) \, .
\end{eqnarray}
Here the superscript symbol on the transition matrix signifies the basis choice.  These transition operators satisfy the required completeness relation and can be used to calculate the corresponding value of $S$.  In doing so, we find $S = 3$, beyond what is possible with the quantum machines, which is in agreement with the classical limit mentioned in \cite{guhne}. We note that such machines are rare, as demonstrated by our sampling and as discussed in the previous paragraph, but still exist nonetheless. These numerical results illustrate that the classical and quantum bounds of $S$ in case of temporal CHSH scenarios can get violated by classical stochastic processes that are describable by macrorealistic theories and noncontextual models \cite{LG,guhne}. Moreover, quantum systems achieve the Tsirelson bound only when projective-like operators are considered for measurements. Given the results, a detailed analysis of the corresponding models from the perspective of quantum foundations is beyond the scope of this paper, however, the obtained results emphasise that CHSH scores lack the adequacy in characterising temporal processes.

The reason for this is that the formulation of the classical machines is actually more general than that of the quantum machines.  In particular, the classical machines do not have the requirement that they generate an orthonormal basis, as can be seen in Eqs.~(\ref{Eq:Example_MC_Alice}) and (\ref{Eq:Example_MC_Bob}), and thus can possess stronger correlations because of this.  As such, the total value of $S$ is not necessarily an appropriate measure of temporal correlations between quantum and classical machines.

\subsection{Longevity of correlations in quantum and classical machines}

Instead of comparing the CHSH score in the case of Alice and Bob's measurements immediately following each other, we can account for some delay between their measurements where the system is perturbed.  This is done by introducing a tertiary evolution channel corresponding to Charlie, with transition operators or Kraus operators $C^{(i)}$.  Separating Alice and Bob's measurements allows us to determine the range of their correlations.  We therefore repeat the sampling procedure as before, but now add an extra evolution according to a full Markovian mapping of Charlie's evolution between.  In each sampling, Charlie chooses his basis of operators and applies the Markovian evolution $t$ times.  For larger values of $t$, we should find that Alice and Bob's outcomes become independent of each other.  Our delayed-time probabilities then become
\begin{eqnarray}
	p(ij,t|A_n B_m) & = & (1,1) B_m^{(j)} (C^{(-1)} + C^{(+1)})^t A_n^{(i)} \eta ~~~
\end{eqnarray}
for the classical case, whilst for the quantum case we utilise standard Markovian maps over the potential evolution pathways.  This means now using a density matrix approach and finding
\begin{eqnarray}
	p(ij,t|A_n B_m) &= & \mathrm{Tr} \left(K_B^{(j)} \mathcal{T}^t K_A^{(i)} \rho {K_A^{(i)}}^\dagger {K_B^{(j)}}^\dagger \right), ~~
\end{eqnarray}
where applies the general Markovian map over Charlie's measurements $t$ times.  Allowing for such a disruptive evolution of Charlie, we now see modified results for the sampled CHSH scores, as shown in Fig.~\ref{Fig:CHSH_time_lag}.  In particular, we see that the quantum versions better maintain their correlations than in the classical version.  This means that although the classical machines are capable of obtaining different scores, due to their differing probability structures, they are fragile and cannot maintain their correlations as well as quantum processes.

\section{Conclusions}\label{sec6}

In this paper, we have analysed temporal correlations of quantum and classical processes and have shown that some standard correlation measures, such as the CHSH inequality, are inappropriate when applied to temporal processes. In doing so, we have demonstrated that the most general classical processes have different probability structures than their quantum counterparts.  As a result, in particular, we see that classical stochastic processes, which can be described by macrorealistic theories and noncontextual models, can surpass the standard classical limit of $S=2$ in these scenarios.  But even stronger than this, we have seen that they can even surpass the limit of $S=2\sqrt{2}$, with a simple example given with $S=3$. Such processes are rare, which is demonstrated by the results of our sampling procedures.  In general, the classical and quantum machines behave similarly in terms of their CHSH score when projective-like Kraus operators are considered for measurement, although restricting to a subset of quantum measurements corresponding to projectors significantly enhances the frequency of generating machines with scores above 2. 

Despite this, we demonstrate that one advantage of quantum temporal processes over classical temporal ones comes from their ability to maintain long-time correlations. While the classical Hidden Markov Models that we consider in this paper can only assume one of two possible hidden states, the analogous Hidden Quantum Markov Models can be prepared in a superposition of two orthogonal quantum states with the respective complex coefficients depending in a much more complex way on the history of the machine. In this paper, this feature is illustrated with an example where the machines experience a random channel between measurements which scrambles the information between Alice and Bob. When considering this case, we see, particularly in the projective case, that the CHSH score is preserved somewhat in the quantum case even under multiple loss operations, while this is not the case for classical machines.  This behavior can indeed be understood by the presence of long-range correlations in quantum systems \cite{AlRabsi2024}. We therefore conclude from this that looking at long range correlations, as opposed to immediate sequential effects, is a more useful metric for characterising temporal processes in the guise of quantum enhancements.

Indeed, when considering only a single time step without scrambling, it appears that classical processes can be more powerful than the quantum ones, since they are able to obtain a higher CHSH score. This, however, is potentially misleading due to the differing probability structure of the classical process and can lead to the assertion that the classical machine can outperform the quantum one in some sense. Instead, our investigation suggests that CHSH scores are often an inappropriate quantitative measure for temporal processes when the correlations are considered instantaneously.  Instead, it is better to quantify them through alternative temporal conditions, such as through their long-range properties. To characterise them fully and to be able to distinguish classical from quantum, we need to consider other measures, such as the {\it out-of-time-ordered correlations} (OTOC) formalism \cite{otoc,OTOC_markov,otoc_qm,mi_otoc,otoc} or the probabilities for the occurrence of certain words, as illustrated in Ref.~\cite{AlRabsi2024}. Understanding the different types of stochastic processes is key for identifying useful quantum resources and for utilising them in quantum processes with quantum technology applications, such as stochastic-number representations and fault-tolerant stochastic computation \cite{chen}. As our paper illustrates, non-classical temporal quantum correlations that are easier to realise than quantum entanglement can lead to a quantum enhancement but only if the right scenarios are considered.

\vspace*{0.5cm}
\noindent {\it Acknowledgements.}
This research is supported by the National Research Foundation, Singapore, under its Competitive Research Programme  (CRP30-2023-0033) and A*STAR under its Quantum Innovation Centre (Q.InC) Strategic Research and Translational Thrust. Any opinions, findings and conclusions or recommendations expressed in this material are those of the author(s) and do not reflect the views of National Research Foundation, Singapore and A*STAR.

\bibliography{reference}

@article{BB84,
title = {Quantum cryptography: Public key distribution and coin tossing},
journal = {Theoretical Computer Science},
volume = {560},
pages = {7-11},
year = {2014},
note = {Theoretical Aspects of Quantum Cryptography – celebrating 30 years of BB84},
issn = {0304-3975},
doi = {https://doi.org/10.1016/j.tcs.2014.05.025},
url = {https://www.sciencedirect.com/science/article/pii/S0304397514004241},
author = {Charles H. Bennett and Gilles Brassard}
}

@article{E91,
  title = {Quantum cryptography based on Bell's theorem},
  author = {Ekert, Artur K.},
  journal = {Phys. Rev. Lett.},
  volume = {67},
  issue = {6},
  pages = {661--663},
  numpages = {0},
  year = {1991},
  month = {Aug},
  publisher = {American Physical Society},
  doi = {10.1103/PhysRevLett.67.661},
  url = {https://link.aps.org/doi/10.1103/PhysRevLett.67.661}
}

@article{tele,
  title = {Teleporting an unknown quantum state via dual classical and Einstein-Podolsky-Rosen channels},
  author = {Bennett, Charles H. and Brassard, Gilles and Crepeau, Claude and Jozsa, Richard and Peres, Asher and Wootters, William K.},
  journal = {Phys. Rev. Lett.},
  volume = {70},
  issue = {13},
  pages = {1895--1899},
  numpages = {0},
  year = {1993},
  month = {Mar},
  publisher = {American Physical Society},
  doi = {10.1103/PhysRevLett.70.1895},
  url = {https://link.aps.org/doi/10.1103/PhysRevLett.70.1895}
}

@article{densecoding,
  title = {Communication via one- and two-particle operators on Einstein-Podolsky-Rosen states},
  author = {Bennett, Charles H. and Wiesner, Stephen J.},
  journal = {Phys. Rev. Lett.},
  volume = {69},
  issue = {20},
  pages = {2881--2884},
  numpages = {0},
  year = {1992},
  month = {Nov},
  publisher = {American Physical Society},
  doi = {10.1103/PhysRevLett.69.2881},
  url = {https://link.aps.org/doi/10.1103/PhysRevLett.69.2881}
}

@article{aspect,
  title = {Experimental Test of Bell's Inequalities Using Time-Varying Analyzers},
  author = {Aspect, Alain and Dalibard, Jean and Roger, Gerard},
  journal = {Phys. Rev. Lett.},
  volume = {49},
  issue = {25},
  pages = {1804--1807},
  numpages = {0},
  year = {1982},
  month = {Dec},
  publisher = {American Physical Society},
  doi = {10.1103/PhysRevLett.49.1804},
  url = {https://link.aps.org/doi/10.1103/PhysRevLett.49.1804}
}

@article{EPR,
  title = {Can Quantum-Mechanical Description of Physical Reality Be Considered Complete?},
  author = {Einstein, A. and Podolsky, B. and Rosen, N.},
  journal = {Phys. Rev.},
  volume = {47},
  issue = {10},
  pages = {777--780},
  numpages = {0},
  year = {1935},
  month = {May},
  publisher = {American Physical Society},
  doi = {10.1103/PhysRev.47.777},
  url = {https://link.aps.org/doi/10.1103/PhysRev.47.777}
}

@article{hardy,
  title = {Nonlocality for two particles without inequalities for almost all entangled states},
  author = {Hardy, Lucien},
  journal = {Phys. Rev. Lett.},
  volume = {71},
  issue = {11},
  pages = {1665--1668},
  numpages = {0},
  year = {1993},
  month = {Sep},
  publisher = {American Physical Society},
  doi = {10.1103/PhysRevLett.71.1665},
  url = {https://link.aps.org/doi/10.1103/PhysRevLett.71.1665}
}

@article{bell,
  title = {On the Einstein Podolsky Rosen paradox},
  author = {Bell, J. S.},
  journal = {Physics Physique Fizika},
  volume = {1},
  issue = {3},
  pages = {195--200},
  numpages = {6},
  year = {1964},
  month = {Nov},
  publisher = {American Physical Society},
  doi = {10.1103/PhysicsPhysiqueFizika.1.195},
  url = {https://link.aps.org/doi/10.1103/PhysicsPhysiqueFizika.1.195}
}

@article{chsh,
  title = {Proposed Experiment to Test Local Hidden-Variable Theories},
  author = {Clauser, John F. and Horne, Michael A. and Shimony, Abner and Holt, Richard A.},
  journal = {Phys. Rev. Lett.},
  volume = {23},
  issue = {15},
  pages = {880--884},
  numpages = {0},
  year = {1969},
  month = {Oct},
  publisher = {American Physical Society},
  doi = {10.1103/PhysRevLett.23.880},
  url = {https://link.aps.org/doi/10.1103/PhysRevLett.23.880}
}

@article{tsirelson,
  title = {Quantum analogues of the Bell inequalities. The case of two spatially separated domains},
  author = {B. S. Tsirelson},
  journal = {Journal of Soviet Mathematics},
  year = {1987},
  volume = {36},
  number = {4},
  pages = {557--570},
  isbn = {1573-8795},
  doi = {10.1007/BF01663472},
  url = {https://doi.org/10.1007/BF01663472}
}

@article{kochen,
  title = {The Problem of Hidden Variables in Quantum Mechanics},
  author = {Kochen, S. and Specker, E. P.},
  journal = {Journal of Mathematics and Mechanics},
  publisher = {Indiana University Mathematics Department},
  year = {1967},
  volume = {17},
  number = {1},
  pages = {59--87},
  isbn = {00959057, 19435274},
  url = {http://www.jstor.org/stable/24902153}
}

@article{mermin,
  title = {Hidden variables and the two theorems of John Bell},
  author = {Mermin, N. David},
  journal = {Rev. Mod. Phys.},
  volume = {65},
  issue = {3},
  pages = {803--815},
  numpages = {0},
  year = {1993},
  month = {Jul},
  publisher = {American Physical Society},
  doi = {10.1103/RevModPhys.65.803},
  url = {https://link.aps.org/doi/10.1103/RevModPhys.65.803}
}

@article{bohm,
  title = {A Suggested Interpretation of the Quantum Theory in Terms of "Hidden" Variables. I},
  author = {Bohm, David},
  journal = {Phys. Rev.},
  volume = {85},
  issue = {2},
  pages = {166--179},
  numpages = {0},
  year = {1952},
  month = {Jan},
  publisher = {American Physical Society},
  doi = {10.1103/PhysRev.85.166},
  url = {https://link.aps.org/doi/10.1103/PhysRev.85.166}
}

@article{bell_multi,
doi = {10.1088/1367-2630/aaff2d},
url = {https://doi.org/10.1088/1367-2630/aaff2d},
year = {2019},
month = {feb},
publisher = {IOP Publishing},
volume = {21},
number = {2},
pages = {023016},
author = {Curchod, Florian J and Almeida, Mafalda L and Acín, Antonio},
title = {A versatile construction of Bell inequalities for the multipartite scenario},
journal = {New Journal of Physics}
}

@article{horodecki_ent,
  title = {Quantum entanglement},
  author = {Horodecki, Ryszard and Horodecki, Pawel and Horodecki, Michal and Horodecki, Karol},
  journal = {Rev. Mod. Phys.},
  volume = {81},
  issue = {2},
  pages = {865--942},
  numpages = {0},
  year = {2009},
  month = {Jun},
  publisher = {American Physical Society},
  doi = {10.1103/RevModPhys.81.865},
  url = {https://link.aps.org/doi/10.1103/RevModPhys.81.865}
}

@article{Schrodinger, 
title={Discussion of Probability Relations between Separated Systems}, 
volume={31}, 
DOI={10.1017/S0305004100013554}, 
number={4}, 
journal={Mathematical Proceedings of the Cambridge Philosophical Society}, 
author={Schrödinger, E.}, 
year={1935}, 
pages={555–563}}

@article{zurek,
  title = {Quantum Discord: A Measure of the Quantumness of Correlations},
  author = {Ollivier, Harold and Zurek, Wojciech H.},
  journal = {Phys. Rev. Lett.},
  volume = {88},
  issue = {1},
  pages = {017901},
  numpages = {4},
  year = {2001},
  month = {Dec},
  publisher = {American Physical Society},
  doi = {10.1103/PhysRevLett.88.017901},
  url = {https://link.aps.org/doi/10.1103/PhysRevLett.88.017901}
}

@article{vedral,
doi = {10.1088/0305-4470/34/35/315},
url = {https://doi.org/10.1088/0305-4470/34/35/315},
year = {2001},
month = {aug},
publisher = {},
volume = {34},
number = {35},
pages = {6899},
author = {L Henderson and V Vedral},
title = {Classical, quantum and total correlations},
journal = {Journal of Physics A: Mathematical and General}
}

@article{mid,
  title = {Using measurement-induced disturbance to characterize correlations as classical or quantum},
  author = {Luo, Shunlong},
  journal = {Phys. Rev. A},
  volume = {77},
  issue = {2},
  pages = {022301},
  numpages = {5},
  year = {2008},
  month = {Feb},
  publisher = {American Physical Society},
  doi = {10.1103/PhysRevA.77.022301},
  url = {https://link.aps.org/doi/10.1103/PhysRevA.77.022301}
}

@article{LG,
  title = {Quantum mechanics versus macroscopic realism: Is the flux there when nobody looks?},
  author = {Leggett, A. J. and Garg, Anupam},
  journal = {Phys. Rev. Lett.},
  volume = {54},
  issue = {9},
  pages = {857--860},
  numpages = {0},
  year = {1985},
  month = {Mar},
  publisher = {American Physical Society},
  doi = {10.1103/PhysRevLett.54.857},
  url = {https://link.aps.org/doi/10.1103/PhysRevLett.54.857}
}

@article{L,
doi = {10.1088/0953-8984/14/15/201},
url = {https://doi.org/10.1088/0953-8984/14/15/201},
year = {2002},
month = {apr},
publisher = {},
volume = {14},
number = {15},
pages = {R415},
author = {A J Leggett},
title = {Testing the limits of quantum
mechanics:  motivation, state of play, prospects},
journal = {Journal of Physics: Condensed Matter}
}

@article{guhne,
  title = {Bounding Temporal Quantum Correlations},
  author = {Budroni, Costantino and Moroder, Tobias and Kleinmann, Matthias and Guhne, Otfried},
  journal = {Phys. Rev. Lett.},
  volume = {111},
  issue = {2},
  pages = {020403},
  numpages = {5},
  year = {2013},
  month = {Jul},
  publisher = {American Physical Society},
  doi = {10.1103/PhysRevLett.111.020403},
  url = {https://link.aps.org/doi/10.1103/PhysRevLett.111.020403}
}

@article{Fritz,
doi = {10.1088/1367-2630/12/8/083055},
url = {https://doi.org/10.1088/1367-2630/12/8/083055},
year = {2010},
month = {aug},
publisher = {},
volume = {12},
number = {8},
pages = {083055},
author = {Fritz, Tobias},
title = {Quantum correlations in the temporal Clauser–Horne–Shimony–Holt (CHSH) scenario},
journal = {New Journal of Physics},
}

@article{Hoffmann,
doi = {10.1088/1367-2630/aae87f},
url = {https://doi.org/10.1088/1367-2630/aae87f},
year = {2018},
month = {oct},
publisher = {IOP Publishing},
volume = {20},
number = {10},
pages = {102001},
author = {Hoffmann, Jannik and Spee, Cornelia and Gühne, Otfried and Budroni, Costantino},
title = {Structure of temporal correlations of a qubit},
journal = {New Journal of Physics},
}

@article{multi_LG,
  title = {Multiple-measurement Leggett-Garg inequalities},
  author = {Barbieri, Marco},
  journal = {Phys. Rev. A},
  volume = {80},
  issue = {3},
  pages = {034102},
  numpages = {3},
  year = {2009},
  month = {Sep},
  publisher = {American Physical Society},
  doi = {10.1103/PhysRevA.80.034102},
  url = {https://link.aps.org/doi/10.1103/PhysRevA.80.034102}
}

@article{bell_povm,
title = {Bell's inequality without alternative settings},
journal = {Physics Letters A},
volume = {313},
number = {1},
pages = {1-7},
year = {2003},
issn = {0375-9601},
doi = {https://doi.org/10.1016/S0375-9601(03)00722-9},
url = {https://www.sciencedirect.com/science/article/pii/S0375960103007229},
author = {Adán Cabello}
}

@article{IS1,
  title = {Information scrambling at finite temperature in local quantum systems},
  author = {Sahu, Subhayan and Swingle, Brian},
  journal = {Phys. Rev. B},
  volume = {102},
  issue = {18},
  pages = {184303},
  numpages = {22},
  year = {2020},
  month = {Nov},
  publisher = {American Physical Society},
  doi = {10.1103/PhysRevB.102.184303},
  url = {https://link.aps.org/doi/10.1103/PhysRevB.102.184303}
}

@article{IS2,
  title = {Butterfly effect in interacting Aubry-Andre model: Thermalization, slow scrambling, and many-body localization},
  author = {Xu, Shenglong and Li, Xiao and Hsu, Yi-Ting and Swingle, Brian and Das Sarma, S.},
  journal = {Phys. Rev. Res.},
  volume = {1},
  issue = {3},
  pages = {032039},
  numpages = {6},
  year = {2019},
  month = {Dec},
  publisher = {American Physical Society},
  doi = {10.1103/PhysRevResearch.1.032039},
  url = {https://link.aps.org/doi/10.1103/PhysRevResearch.1.032039}
}

@article{qchaos,
  title = {Information Scrambling and Loschmidt Echo},
  author = {Yan, Bin and Cincio, Lukasz and Zurek, Wojciech H.},
  journal = {Phys. Rev. Lett.},
  volume = {124},
  issue = {16},
  pages = {160603},
  numpages = {6},
  year = {2020},
  month = {Apr},
  publisher = {American Physical Society},
  doi = {10.1103/PhysRevLett.124.160603},
  url = {https://link.aps.org/doi/10.1103/PhysRevLett.124.160603}
}

@article{bhole,
  title = {Stringy effects in scrambling},
  author = {Stephen H. Shenker and Douglas Stanford},
  journal = {Journal of High Energy Physics},
  year = {2015},
  volume = {2015},
  number = {5},
  pages = {132},
  isbn = {1029-8479},
  doi = {10.1007/JHEP05(2015)132},
  url = {https://doi.org/10.1007/JHEP05(2015)132}
}

@article{otoc,
   author = {Larkin, A.I. and Ovchinnikov, Yu. N.},
   title = "{Quasiclassical Method in the Theory of Superconductivity}",
   journal = {Soviet Journal of Experimental and Theoretical Physics},
   year = 1969,
   month = jun,
   volume = {28},
   pages = {1200},
   adsurl = {https://ui.adsabs.harvard.edu/abs/1969JETP...28.1200L}
}

@article{susskind,
doi = {10.1088/1126-6708/2008/10/065},
url = {https://doi.org/10.1088/1126-6708/2008/10/065},
year = {2008},
month = {oct},
publisher = {},
volume = {2008},
number = {10},
pages = {065},
author = {Yasuhiro Sekino and L. Susskind},
title = {Fast scramblers},
journal = {Journal of High Energy Physics},
}

@article{chaos,
  title = {Chaos Signatures in the Short and Long Time Behavior of the Out-of-Time Ordered Correlator},
  author = {Garcia-Mata, Ignacio and Saraceno, Marcos and Jalabert, Rodolfo A. and Roncaglia, Augusto J. and Wisniacki, Diego A.},
  journal = {Phys. Rev. Lett.},
  volume = {121},
  issue = {21},
  pages = {210601},
  numpages = {5},
  year = {2018},
  month = {Nov},
  publisher = {American Physical Society},
  doi = {10.1103/PhysRevLett.121.210601},
  url = {https://link.aps.org/doi/10.1103/PhysRevLett.121.210601}
}

@article{mi_otoc,
doi = {10.1088/2058-9565/ab8ebb},
url = {https://doi.org/10.1088/2058-9565/ab8ebb},
year = {2020},
month = {may},
publisher = {IOP Publishing},
volume = {5},
number = {3},
pages = {035005},
author = {Touil, Akram and Deffner, Sebastian},
title = {Quantum scrambling and the growth of mutual information},
journal = {Quantum Science and Technology}
}

@article{otoc_qm,
  title = {Out-of-time-order correlators in quantum mechanics},
  author = {Koji Hashimoto and Keiju Murata and Ryosuke Yoshii},
  journal = {Journal of High Energy Physics},
  year = {2017},
  volume = {2017},
  number = {10},
  pages = {138},
  isbn = {1029-8479},
  doi = {10.1007/JHEP10(2017)138},
  url = {https://doi.org/10.1007/JHEP10(2017)138}
}

@article{OTOC_markov,
  title = {Two-step phantom relaxation of out-of-time-ordered correlations in random circuits},
  author = {Bensa, J.c.v. and Znidaric, M},
  journal = {Phys. Rev. Res.},
  volume = {4},
  issue = {1},
  pages = {013228},
  numpages = {16},
  year = {2022},
  month = {Mar},
  publisher = {American Physical Society},
  doi = {10.1103/PhysRevResearch.4.013228},
  url = {https://link.aps.org/doi/10.1103/PhysRevResearch.4.013228}
}

@article{temporal_steer,
  title = {Experimental temporal quantum steering},
  author = {Karol Bartkiewicz and Antonín Černoch and Karel Lemr and Adam Miranowicz and Franco Nori},
  journal = {Scientific Reports},
  year = {2016},
  volume = {6},
  number = {1},
  pages = {38076},
  isbn = {2045-2322},
  doi = {10.1038/srep38076},
  url = {https://doi.org/10.1038/srep38076}
}

@article{PR_box,
  title = {Quantum nonlocality as an axiom},
  author = {Sandu Popescu and Daniel Rohrlich},
  journal = {Foundations of Physics},
  year = {1994},
  volume = {24},
  number = {3},
  pages = {379--385},
  isbn = {1572-9516},
  doi = {10.1007/BF02058098},
  url = {https://doi.org/10.1007/BF02058098}
}

@article{quantum_causality,
  title = {Quantum causality},
  author = {Časlav Brukner},
  journal = {Nature Physics},
  year = {2014},
  volume = {10},
  number = {4},
  pages = {259--263},
  isbn = {1745-2481},
  doi = {10.1038/nphys2930},
  url = {https://doi.org/10.1038/nphys2930}
}

@misc{Garg2025,
      title={Assessing the dynamical assumptions in Tsirelson inequality tests of non-classicality in harmonic oscillators}, 
      author={Arush Garg and Jonathan Halliwell and Taejas Venkataraman},
      year={2025},
      eprint={2509.03166},
      archivePrefix={arXiv},
      primaryClass={quant-ph},
      url={https://arxiv.org/abs/2509.03166}, 
}

@ARTICLE{AlRabsi2024,
AUTHOR={Al Rasbi, Kawthar  and Clark, Lewis A.  and Beige, Almut },
TITLE={Quantum physics cannot be captured by classical linear hidden variable theories even in the absence of entanglement},
JOURNAL={Frontiers in Physics},       
VOLUME={12},  
YEAR={2024},  
URL={https://www.frontiersin.org/journals/physics/articles/10.3389/fphy.2024.1325239},
DOI={10.3389/fphy.2024.1325239},  
ISSN={2296-424X},
}

@book{Nielsen:2012yss,
    author = {Nielsen, M. A. and Chuang, I. L.},
    title = {Quantum Computation and Quantum Information},
    doi = {10.1017/cbo9780511976667},
    isbn = {978-0-521-63503-5},
    publisher = {Cambridge University Press},
    month = {6},
    year = {2012}
}

@book{q_channel,
  title = {States, Effects, and Operations: Fundamental Notions of Quantum Theory},
  author    = {Kraus, K and Bohm, A and Dollard, J D and Wootters, W H},
  publisher = {Springer Berlin Heidelberg},
  year      =  1983
}

@article{DiVincenzo,
author = {DiVincenzo, David P.},
title = {The Physical Implementation of Quantum Computation},
journal = {Fortschritte der Physik},
volume = {48},
number = {9-11},
pages = {771-783},
doi = {https://doi.org/10.1002/1521-3978(200009)48:9/11<771::AID-PROP771>3.0.CO;2-E},
url = {https://onlinelibrary.wiley.com/doi/abs/10.1002/1521-3978%28200009%2948%3A9/11%3C771%3A%3AAID-PROP771%3E3.0.CO%3B2-E},
year = {2000}
}

@article{Lewis1,
  title = {Quantum-enhanced metrology with the single-mode coherent states of an optical cavity inside a quantum feedback loop},
  author = {Clark, Lewis A. and Stokes, Adam and Beige, Almut},
  journal = {Phys. Rev. A},
  volume = {94},
  issue = {2},
  pages = {023840},
  numpages = {12},
  year = {2016},
  month = {Aug},
  publisher = {American Physical Society},
  doi = {10.1103/PhysRevA.94.023840},
  url = {https://link.aps.org/doi/10.1103/PhysRevA.94.023840}
}

@article{Lewis2,
  title = {Quantum jump metrology},
  author = {Clark, Lewis A. and Stokes, Adam and Beige, Almut},
  journal = {Phys. Rev. A},
  volume = {99},
  issue = {2},
  pages = {022102},
  numpages = {12},
  year = {2019},
  month = {Feb},
  publisher = {American Physical Society},
  doi = {10.1103/PhysRevA.99.022102},
  url = {https://link.aps.org/doi/10.1103/PhysRevA.99.022102}
}

@article{Kawthar1,
  title = {Quantum jump metrology in a two-cavity network},
  author = {Rasbi, Kawthar Al and Beige, Almut and Clark, Lewis A.},
  journal = {Phys. Rev. A},
  volume = {106},
  issue = {6},
  pages = {062619},
  numpages = {12},
  year = {2022},
  month = {Dec},
  publisher = {American Physical Society},
  doi = {10.1103/PhysRevA.106.062619},
  url = {https://link.aps.org/doi/10.1103/PhysRevA.106.062619}
}

@article{stinespring,
  title={Positive Functions on C*-Algebras},
  author={Stinespring, W. Forrest},
  journal={Proceedings of the American Mathematical Society},
  volume={6},
  number={2},
  pages={211--216},
  year={1955}
}

@Inbook{Clark2015,
author={Clark, Lewis A. and Huang, Wei and Barlow, Thomas M. and Beige, Almut},
editor={Sanayei, O.Z.I. and  Aliand E. Rossler},
title="Hidden Quantum Markov Models and Open Quantum Systems with Instantaneous Feedback",
bookTitle="ISCS 2014: Interdisciplinary Symposium on Complex Systems",
year="2015",
publisher="Springer International Publishing",
address="Cham",
pages="143--151",isbn="978-3-319-10759-2",doi="10.1007/978-3-319-10759-2_16",url="https://doi.org/10.1007/978-3-319-10759-2_16"}

@article{Monras2011,
    author = "Monras, A. and Beige, A. and Wiesner, K.",
    title = "Hidden Quantum Markov Models and non-adaptive read-out of many-body states",
    journal = "Applied Mathematical and Computational Sciences 3, 93",
    volume = "3",
    pages = "93",
    year = "2011"
}

@article{Rabiner1989,
    author = {Rabiner, L.~R.~},
    title = {A tutorial on hidden Markov models and selected applicationsin speech recognition},
    journal = {Proc.~IEEE},
    volume = {77},
    pages = {257},
    year = {1989}
}

@article{Xue2006,
    author = {Xue, H.},
    title = {Hidden Markov Models Combining Discrete Symbols and Continu-ous Attributes in Handwriting Recognition},
    journal = {IEEE Transactions on PatternAnalysis and Machine Intelligence},
    volume = {28},
    pages = {458},
    year = {2006}
}

@article{Vanluyten2008,
    author = {Vanluyten, B. and Willems, J.~C. and Moor, B.~D.},
    title = {Equivalence of State Repre-sentations for Hidden Markov Models},
    journal = {Systems and Control Letters},
    volume = {57},
    pages = {410},
    year = {2008}
}

@article{Clauser_1978,
doi = {10.1088/0034-4885/41/12/002},
url = {https://doi.org/10.1088/0034-4885/41/12/002},
year = {1978},
month = {dec},
publisher = {},
volume = {41},
number = {12},
pages = {1881},
author = {J F Clauser and A Shimony},
title = {Bell's theorem. Experimental tests and implications},
journal = {Reports on Progress in Physics},
}

@ARTICLE{chen,
  author={Chen, Yuhao and Li, Hongge and Song, Yinjie and Zhu, Xinyu},
  journal={IEEE Transactions on Circuits and Systems I: Regular Papers}, 
  title={Recoding Hybrid Stochastic Numbers for Preventing Bit Width Accumulation and Fault Tolerance}, 
  year={2025},
  volume={72},
  number={3},
  pages={1243-1255},
  doi={10.1109/TCSI.2024.3492054}}
\end{document}